# GRANULAR FLOW MODELLED BY BROWNIAN PARTICLES

T. RIETHMÜLLER, D. ROSENKRANZ, L. SCHIMANSKY-GEIER
*Lehrstuhl für stochastische Physik* [†]
*Institut für Physik, Humboldt–Universität zu Berlin*
*Invalidenstr. 110, 10115 Berlin, Germany*

Granular flows trough pipes show interesting phenomena, e.g. clogging and density waves, 1/f–noise [4,5,6]. These things are fairly good studied by computer–experiments, but there is a lack in theoretical and analytical consideration.
We introduce a simple "minimal" model describing such a flow of granular particles and examine the stability of an initially homogeneous system against perturbations. In order to define the collisions between the granular particles we use two different approaches. For both, the simple and the more advanced collision definition, the model shows the qualitative same behaviour.

## 1 The Model

We consider the motion of granular particles, falling through a thin long pipe, as an one–dimensional motion and describe it with a system of Langevin–Equations:

$$\begin{aligned} \dot{x}_i &= v_i & (i=1\cdots N) \\ \dot{v}_i &= g - \gamma v_i + \sqrt{2\sigma\gamma}\xi_i(t) & (\xi_i(t) - \text{gaussian, white noise}) \end{aligned} \quad (1)$$

The particles are accelerated by the gravity $g$ and interact with the wall and possibly with the air in the pipe. We describe these influences with an effective friction and with stochastic fluctuations of the velocities. In addition to this we have to take into account **collisions** between the Brownian Particles.
The One–particle–distribution–function $P(x,v,t)$ is then defined by the following Boltzmann–Equation:

$$\frac{\partial}{\partial t}P + \frac{\partial}{\partial x}[vP] + \frac{\partial}{\partial v}[(g-\gamma v)P] = \sigma\gamma\frac{\partial^2}{\partial v^2}P + \left(\frac{\partial}{\partial t}P\right)_{Coll.} . \quad (2)$$

Here, $\left(\frac{\partial}{\partial t}P\right)_{Coll.}$ denotes the contribution from the collisions.
Now we can derive the hydrodynamic equations by multiplying the Boltzmann–Equation with $v^k$ and integrating over $v$:

$$\begin{aligned} 0 &= \frac{\partial}{\partial t}\rho + u\frac{\partial}{\partial x}\rho + \rho\frac{\partial}{\partial x}u & (3) \\ 0 &= \frac{\partial}{\partial t}u + u\frac{\partial}{\partial x}u + \frac{k_B}{m\rho}\frac{\partial}{\partial x}(\rho T) - (g-\gamma u) - \frac{1}{\rho}\int dv\, v\left(\frac{\partial}{\partial t}P\right)_{Coll.} & (4) \\ 0 &= \frac{\partial}{\partial t}T + u\frac{\partial}{\partial x}T + 2T\frac{\partial}{\partial x}u + 2\gamma T - 2\frac{m}{k_B}\gamma\sigma + \\ &\quad +2u\frac{m}{k_B\rho}\int dv\, v\left(\frac{\partial}{\partial t}P\right)_{Coll.} - \frac{m}{k_B\rho}\int dv\, v^2\left(\frac{\partial}{\partial t}P\right)_{Coll.} . & (5) \end{aligned}$$

---

[†] http://summa.physik.hu-berlin.de/

In order to get a closed equation system, we assume, that the moments $\overline{(v-u)^k}$ are negligible small for $k \geq 3$.

## 2 Stability of the Granular System

### 2.1 A Simple Collision model

This simple model for binary collisions was introduced by Herman and Prigogine[1] to describe the phenomena of traffic flows. Because of the similarities between traffic and granular flows we use this definition as our first approach.

The model is a so called "Follow the Leader"–model, that means, if a particle with the velocity $v'$ collides with a particle with a lower velocity $v$, both particles move with the velocity $v$ after the collision process.

$$\bigcirc \xrightarrow{v'} \bigcirc \xrightarrow{v} \quad \Rightarrow \quad \bigcirc\bigcirc \xrightarrow{v}$$

For this process one gets the following collision terms:

$$\begin{aligned}
\left(\frac{\partial}{\partial t}P(x,v,t)\right)_{Coll.} &= S_C \int dv'\, P(x,v,t)P(x,v',t)(v'-v) \\
&= S_C P(x,v,t)\rho(x,t)(u(x,t)-v) \quad (6)\\
\int dv \left(\frac{\partial}{\partial t}P\right)_{Coll.} &= 0 \quad (7)\\
\int dv\, v \left(\frac{\partial}{\partial t}P\right)_{Coll.} &= -S_C \frac{k_B}{m}\rho^2 T(x,t) \quad (8)\\
\int dv\, v^2 \left(\frac{\partial}{\partial t}P\right)_{Coll.} &= -2S_C \frac{k_B}{m}u(x,t)\rho^2 T(x,t)\,. \quad (9)
\end{aligned}$$

$S_C$ denotes here the collision cross–section or in other words, the probability, that a collision take place.

Now we can examine the stability of the granular system against perturbations. We make the following perturbation ansatz for the hydrodynamic variables:

$$\begin{aligned}
\rho(x,t) &= \rho_h + \delta\rho &= \rho_h + \delta\rho_0 \exp(-\alpha t + ikx) \\
u(x,t) &= u_h + \delta u &= u_h + \delta u_0 \exp(-\alpha t + ikx) \\
T(x,t) &= T_h + \delta T &= T_h + \delta T_0 \exp(-\alpha t + ikx)\,,
\end{aligned} \quad (10)$$

insert them in the corresponding hydrodynamic equations, drop all quadratic perturbations and get a cubic equation for the value of $\alpha$. In dependence of the parameters of the system and the wavenumber $k$ one finds critical densities $\rho_c$ where the stability of the system changes. That means, that for densities below $\rho_c$ the real part of $\alpha$ is positive, the system is stable against perturbations and vice versa. We assume periodic boundary conditions and so the wavenumber $k$ is discrete: $k = \frac{2\pi}{L}n \quad (n = \pm 1, \pm 2, \cdots)$. Figure 1 shows the dependence of the critical density $\rho_c$ from the mode-number $n$. One sees, that especially short length perturbations are able to destabilise the granular system. This stands in strong contrast to results found for traffic flows[1,2,3], where the long range fluctuations are the critical ones. Our results are not surprising, if one imagine, that a local large gradient of the velocities will lead to a high collision rate at this place. For sufficient high densities this continues avalanche–like and leads to clusters with high local density and small average velocity.

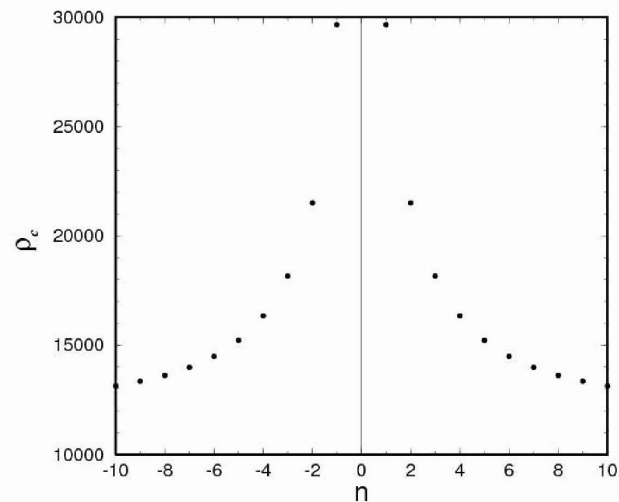

Figure 1: The critical density $\rho_c$ for various mode-numbers $n$

For large wave numbers we get a low limiting critical density of $\rho_c^{min} \approx 12000$.

To check these results we carried out some computer simulations. The pictures in Fig.2 show snapshots from a simulation with a subcritical density ($\rho < \rho_c^{min}$) at t=0s, 3s, 500s. One sees an almost homogeneous density–distribution without any cluster formation. Notice, that the velocity distribution $w(v,t) = \int_0^L P\,dx$ remains gaussian.

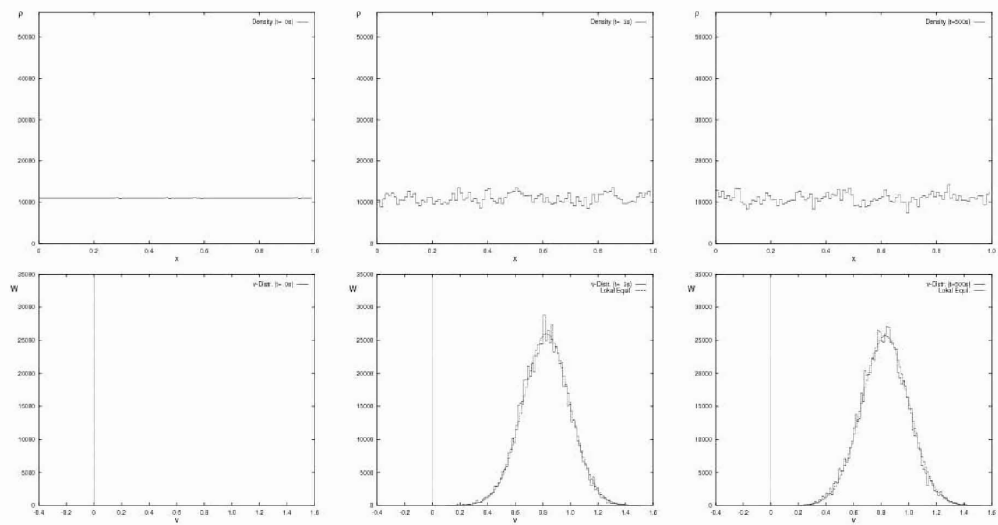

Figure 2: Simulation results in the subcritical regime

The next series of snapshots (Fig.3) at the same times shows the behaviour for a supercritical density. One sees clearly the formation of two moving clusters. A further characteristic is the non gaussian velocity distribution $w(v,t)$. The deviation from the Gaussian results in our opinion from the fact, that one has at least two distinct velocity distributions in the system, one with a high average–velocity (at low density regions) and one with a small average–velocity (at high density regions). The overlayed curves in the distribution plots show the expected shape of such an bimodal distribution. The deviation from the measured distribution is due to the fact, that the influences of the slopes of the clusters are neglected.

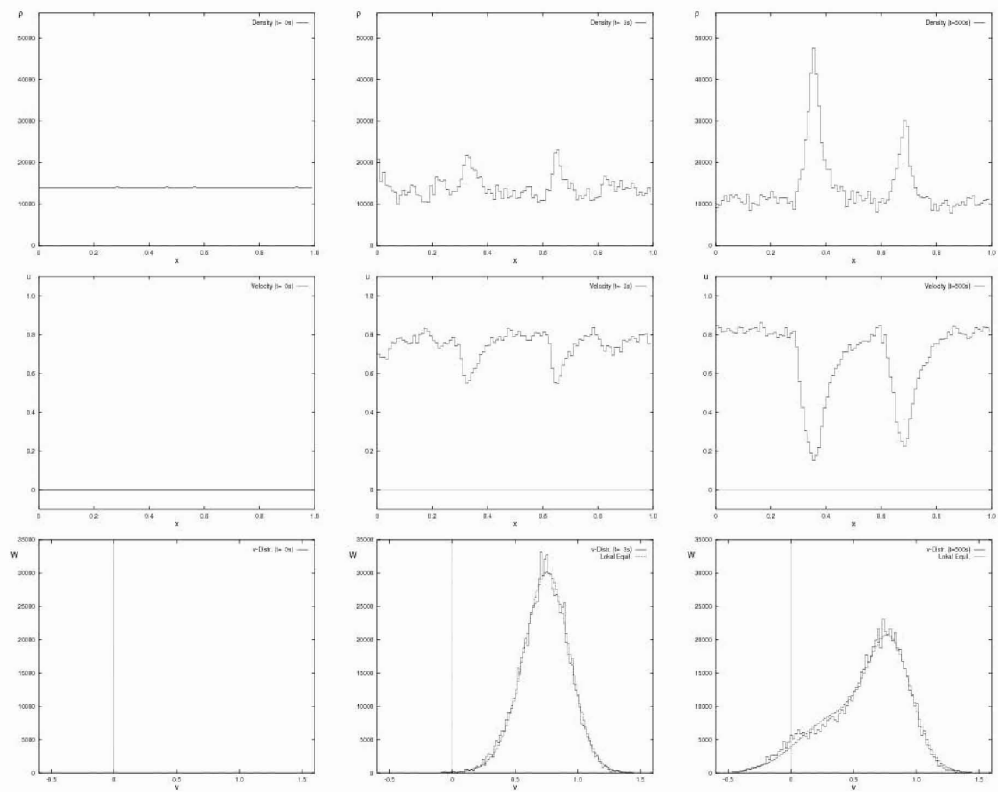

Figure 3: Simulation results in the supercritical regime

## 2.2 A Second Collision Model

One point, one can criticise in the simple model, is, that the total momentum is not preserved during the collision process. Therefore we also considered collisions with momentum preservation. The collision process is defined by the equations:

$$\begin{aligned} (v'_i - v'_j) &= -\epsilon \ (v_i - v_j) \qquad 0 \leq \epsilon \leq 1 \\ (v'_i + v'_j) &= \ (v_i + v_j) \, . \end{aligned} \tag{11}$$

The restitution coefficient $\epsilon$ allows us to change the kind of collisions from totally inelastic ($\epsilon = 0$) to elastic ($\epsilon = 1$). The collision operator reads then as follows:

$$\left( \frac{\partial}{\partial t} P(x, v, t) \right)_{Coll.} = S_C \left( \frac{4}{(1+\epsilon)^2} \int dv' \, P(x, v', t) P(x, \tfrac{2v - (1-\epsilon)v'}{1+\epsilon}, t) |v' - v| \right. \\ \left. - \int dv' \, P(x, v', t) P(x, v, t) |v' - v| \right) . \tag{12}$$

This operator is much more complicated as the former one, but one can show, that the mass and momentum preservation is fulfilled:

$$\int dv \left(\frac{\partial}{\partial t}P\right)_{Coll.} = \int dv\, v \left(\frac{\partial}{\partial t}P\right)_{Coll.} = 0. \qquad (13)$$

Unfortunately, the last term (cf. Eq.(9)) cannot be evaluated in general; one has to assume an ansatz for the distribution function $P(x,v,t)$ to get some results.

If one sets the collision cross-section to zero, there are no collisions and the stationary velocity distribution is a Gaussian around the mean velocity with the width $\sigma = \frac{k_B}{m}T$. For small $S_C$ it should be allowed to use such a Gaussian as a first approach for the velocity distribution. With such an ansatz one get the following result for the integral:

$$\int dv\, v^2 \left(\frac{\partial}{\partial t}P\right)_{Coll.} = -2S_C \frac{1-\epsilon^2}{\sqrt{\pi}} \sqrt{\left(\frac{k_B}{m}T\right)^3}. \qquad (14)$$

After inserting it in the hydrodynamic equations one can proceed with the whole linear stability analysis.

The behaviour with this collision model is similar to that of our first simple approach, one gets the qualitative same dependence of the critical density from the wavenumber. Only short range fluctuations are able to destabilise the system. In order to check these results we currently perform computer simulations.

## 3 Conclusions and Remarks

- We have used a simple model consisting of Brownian Particles with collision interaction to model the granular flow trough a pipe. We considered two different approaches to the collision process and showed, that in both cases sufficiently large densities leads to an unstable state of the flow.

- Computer simulations support these results.

- Strictly speaking, our considerations are only valid in a low density regime. Especially the approach of two–particle–collisions is not sufficient for higher densities. We have to check the influence of such phenomena.

- Does our simple "effective" model really reflect real granular flows? We have to compare our results with results of experiments.